# Switchable bifunctional metasurfaces: nearly perfect retroreflection and absorption at THz regime

SHAN ZHU,[1, 2] YANYAN CAO,[1] YANGYANG FU,[3, 4] XIAOCHAO LI,[2] LEI GAO,[1, 5] HUANYANG CHEN [2] AND YADONG XU[1, 6]

[1]School of Physical Science and Technology, Soochow University, Suzhou 215006, China
[2]School of Electronic Science and Engineering, Xiamen University, Xiamen 361005, China
[3]College of Science, Nanjing University of Aeronautics and Astronautics, Nanjing 211106, China
[4]e-mail:yyfu@nuaa.edu.cn
[5]e-mail:leigao@suda.edu.cn
[6]e-mail:ydxu@suda.edu.cn

**Here we make use of vanadium dioxide ($VO_2$) to design a bifunctional metasurface working at the same targeted frequency. With the increase of temperature, the functionality of the designed metasurface can switch from a multi-channel retroreflector to a perfect absorber, caused by the phase transition of $VO_2$ from insulator to conductor. Different from traditional bifunctional metasurfaces designed by simple composition of two functionalities, our proposed bifunctional metasurface is based on the interaction between two functionalities. The device shows good potential for the combination of wavefront manipulation and optical absorption, therefore providing a promising approach for switchable detection and anti-detection devices.**

Optical metasurfaces [1-5] have drawn considerable attention because they have provided unprecedented abilities to control electromagnetic waves. By carefully engineering the meta-atoms to enhance the interaction between metasurfaces and light, a variety of applications based on metasurfaces have been proposed [6-16], such as optical holograms [6, 7], anomalous reflection [8, 9], metalens [10, 11], parity-dependent perfect anomalous diffraction [14] and so on. So far, most of metasurface-based applications are static and their functionalities highly depend on the designed geometric configuration. However, optical elements with single-function are insufficient for the integration and miniaturization of compact systems in advanced information processing technology.

In recent years, metasurfaces with multifarious functionalities and dynamically switchable property grow into an emerging research area. Great efforts [17-19] have been made to realize multifunctional metasurfaces by incorporating active modulations into the passive structures. For instance, at microwave frequency range, by employing integrated circuits (ICs) [17] to independently adjust resistance and reactance, an intelligent metasurface was proposed, with its functionality that can be switched from perfect absorption to perfect anomalous reflection. At high frequency range, such as terahertz (THz) frequencies, based on phase change properties of vanadium dioxide ($VO_2$), a group [18] numerically designed a multifunctional metasuface, which can switch its functionality from absorber to polarizer. However, to the best of our knowledge, a single metasurface with dual functionalities of both perfect retroreflection and perfect absorber at the same operating frequency, has not reported yet. Retroreflection is a unique phenomenon that incident wave is reflected back by a surface or device to its source direction [20, 21], which is strongly desired in laser tracking [22] or remote sensing [23]. In contrast with retroreflector used for signal detection, a perfect absorber can effectively eliminate the reflected signal, which could be used for anti-detection. Therefore, a single device with switchable functionalities of retroreflector and absorber can actively choose to be visible and invisible, which has significant values in potential applications, such as dunker and aerobat in military scenes.

In this *Letter*, we design and study a bifunctional metadevice in THz regime by integrating $VO_2$ into a phase-gradient metasurface (PGM) that only consists of two unit cells, which can realize switchable functionalities of nearly perfect retroreflection and perfect absorption at the same target frequency. $VO_2$ is temperature-dependent phase transition material, with the transition temperature about 68℃ [24]. At low temperature, $VO_2$ behaves like a dielectric with low dissipation; when the temperature is beyond the phase change temperature, $VO_2$ is a highly lossy metal. Unlike other multifunctional metasurfaces, our method is not only based on the simple superposition of two single functional metasurfaces, but also considers the coupling between them. It is found that at low temperature, the PGM produces nearly perfect retroreflection. At high temperature, the strong resonance of EM field inside $VO_2$ and the interaction between $VO_2$ and the PGM lead to nearly perfect absorption. Numerical simulations are performed to demonstrate the switchable functionalities of retroreflection and absorption.

Figure 1(a) shows schematically the unit cell of considered PGM, a three-layered structure on Si substrate. The bottom layer is gold

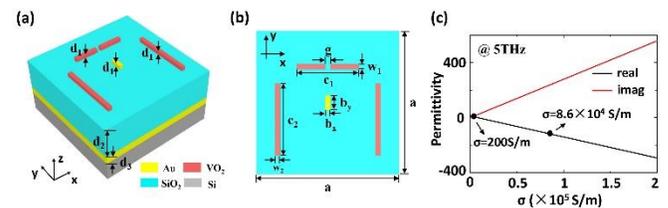

Fig. 1. (a) The unit cell designed for the binary metasurface. (b) The top view of the unit cell. (c) The real and imaginary part of the permittivity of $VO_2$ vs its conductivity at 5.0 THz.

with its optical characteristic described by $\varepsilon_m = 1 - f_{p1}^2 / (f^2 + i\gamma_1 f)$, where $f_{p1} = 2181.53\,THz$ and

$\gamma_1 = 19.11\,THz$ [25]. The middle layer is SiO$_2$ and its permittivity is 3.9 with loss ignored [26]. The top layer consists of a Au stripe, two VO$_2$ stripes and a VO$_2$ bar with gap, with identical thickness. The two VO$_2$ stripes are parallel with the Au stripe along $y$ direction and the VO$_2$ bar is along x direction. Fig. 1(b) depicts the top view of the unit cell. The relative permittivity of VO$_2$ is modeled by Drude model [18], i.e.,

$$\varepsilon(f) = \varepsilon_\infty - f_{p2}^2 \sigma / \sigma_0(f^2 + i\gamma_2 f), \quad (1)$$

where $\varepsilon_\infty = 12$, $f_{p2} = 222.93\,THz$, $\gamma_2 = 9.61\,THz$, $\sigma_0 = 3 \times 10^5\,S/m$ and $\sigma$ is the conductivity of VO$_2$. Here the operating frequency is considered at 5.0 THz. The real and imaginary part of permittivity of VO$_2$ at 5.0 THz are shown in Fig. 1(c). At the low temperature, e.g., $\sigma = 200\,S/m$, due to low loss and deep sub-wavelength thickness ($\lambda/60$), compared with high reflectivity of the Au stripe, the reflectivity of VO$_2$ is almost ignored. In this case, the size of the Au stripe is critical for modulating the reflected phase of the unit cell. When the temperature is beyond the phase change temperature, VO$_2$ experiences an insulator-to-metal transition. At the high temperature, e.g., $\sigma = 8.6 \times 10^4\,S/m$, VO$_2$ is a metal with high loss, which could be potentially used to design optical absorber.

We will first design a binary metasurface at 5.0 THz to achieve three-channel retroreflector [27, 28] at low temperature. As displayed in Fig. 2(a), we construct a supercell consisting of two unit cells, where the Au stripes are designed with different lengths to achieve a phase difference of $\pi$. The dielectric layer in our metasurface works as a substrate for the growth of Au and VO2. When the thickness of the dielectric layer is changed, the length of the Au stripes will be adjusted to satisfy the required phase difference of the two cells. A sketch map of a perfect retroreflector is shown in Fig. 2(b). According to the wave diffraction in phase-gradient metasurfaces [29], the incident and reflected waves are governed by,

$$k_0 \sin\theta_i = k_0 \sin\theta_r + nG, \quad (2)$$

where $G = 2\pi/p$ is reciprocal lattice vector, $k_0 = 2\pi/\lambda$ is wave vector in vacuum space, $n$ represents diffraction order, $\theta_i$ and $\theta_r$ are incident and reflected angle, respectively. The geometrical parameters of the unit cell are optimized as $d_1 = 1.0\,\mu m$, $d_2 = 8.0\,\mu m$, $d_3 = 2.0\,\mu m$, $b_x = 1\,\mu m$, $w_1 = w_2 = 1.0\,\mu m$, $c_1 = 13\,\mu m$, $c_2 = 15\,\mu m$, $g = 1.0\,\mu m$, respectively. As $p = \lambda = 60\mu m$, $G = k_0$ lead to effective diffraction order belonging to $n \in [-1,1]$ in Eq. (2). To realize retroreflection, i.e., $\theta_r = -\theta_i$, Eq. (2) reduces to $2\sin\theta_i = n$, which means the retroreflection could be obtained at $\theta_i = -30°$, $\theta_i = 0°$ and $\theta_i = 30°$ via the diffraction orders of $n = -1$, $n = 0$ and $n = 1$, respectively. In order to obtain high-efficiency retroreflection at oblique incidences, the specular reflection need to be suppressed. For this aim, phase difference of $\pi$ is required for the two unit cells at $\theta_i = -30°$ or $\theta_i = 30°$. The physical mechanism behind is that when the reflected phase difference of the two unit cells is $\pi$, destructive interference of the zero-order diffraction will happen, i.e., the zero-order reflected waves from binary unit cell cancel each other, leading to the suppression of specular reflection.

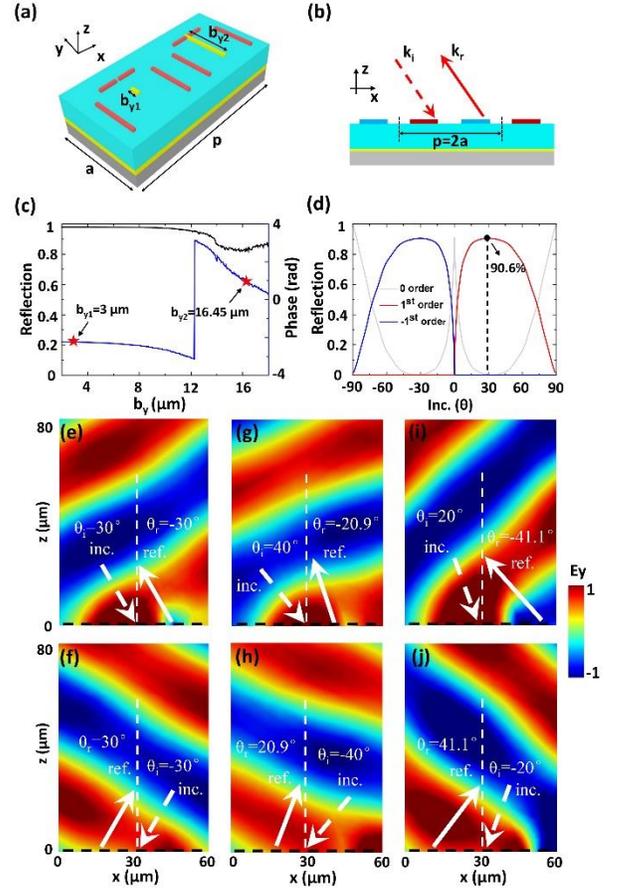

Fig. 2. (a) The schematic of supercell for binary metasurface. (b) A sketch map of perfect retroreflector. (c) The reflection amplitude and phase of the unit cell vs the length of Au stripe. (d) Reflection efficiencies of each diffraction orders. (e)-(f) The reflected field patterns (E$_y$) of different incident angles of $\theta_i = 30°$ (e), $\theta_i = -30°$ (f), $\theta_i = 40°$ (g), $\theta_i = -40°$ (h), $\theta_i = 20°$ (i) and $\theta_i = -20°$ (j).

By varying the length of Au stripe (b$_y$), the reflection amplitude and phase of the unit cell are displayed in Fig. 2(c), where incident wave with $\theta_i = 30°$ bumps on the unit cell with the electric field direction along $y$ direction. It is observed that the length of Au stripe at b$_{y1}$ = 3.0 μm and b$_{y2}$ =16.45 μm, respectively, enables phase difference of $\pi$ for the two unit cells, and their corresponding reflections are 97.9% and 82.3%, with a slight deviation resulted from strong resonance happening in Au stripe at b$_{y2}$=16.45 μm. Fig. 2(d) shows numerically calculated reflection efficiencies of each orders. For $\theta_i = 30°$ ($-30°$), the reflection efficiency of the $n = 1$ ($-1$) order is nearly 90.6%, and other energy of approximately 9.4% is mainly absorbed by the Au stripes. The reflection efficiency of the $n = 0$ order is nearly zero at $\theta_i = \pm 30°$ and 91.1% at $\theta_i = 0°$. In addition, we also find that the reflection efficiency of $n = \pm 1$ order beyond 85.0% can occur near $\theta_i = \pm 30°$, i.e., $\theta_i \in [-50°, -13°]$ and $\theta_i \in [13°, 50°]$, yet with nearly zero reflection in the $n = 0$ order. In these two regions, quasi-retroreflection [30] with high efficiency happens. To clearly see the performance of the retroreflector, the electric field patterns in the x-z plane are presented in Fig. 2(e)-(j). When $\theta_i = \pm 30°$, high-efficiency retroreflected wave is seen (Fig. 2(e) and (f)). For quasi-retroreflection, we select four incident angles for demonstration, i.e., $\theta_i = \pm 20°$ and $\theta_i = \pm 40°$. When $\theta_i = \pm 40°$ (see Fig. 2(g) and (h)), high-efficiency reflected waves are guided to $\theta_r = \mp 20.9°$, which has approximate deviation of $19.1°$ from the exact retroreflection angles. We also observe the phenomenon of quasi-retroreflection with a deviation angle of $21.1°$ for $\theta_i = \pm 20°$ (see Fig. 2(i) and Fig. 2(j)). As shown in Fig. 2, the scattered field patterns are less flat, the reason is that the reflection amplitudes of the two components in the supercell have a little deviation, which makes zero-order diffraction wave radiate out with low efficiency.

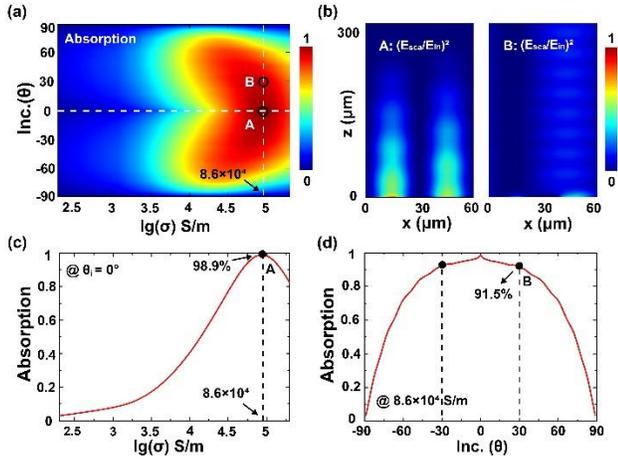

Fig. 3. (a) The absorption efficiency of the metasurface as a function of incident angle and the conductivity of VO$_2$. (b) Field patterns at point A (left) and point B (right). (c) The absorption efficiency for $\theta_i = 0°$. (d) The absorption efficiency for $\sigma = 8.6 \times 10^4 S/m$.

Next we will explore the functionality of perfect absorber in the designed metasurface at high temperature (i.e., VO$_2$ with larger $\sigma$), yet with other parameters unchanged. Figure 3(a) shows the absorption efficiency at 5 THz as a function of incident angle and the conductivity of VO$_2$. The absorption efficiency is defined as $A=1-R$, where $R$ is reflection and the transmitted wave is zero due to the bottom gold layer. When $\theta_i = 0°$, indicated by the horizontal white dashed line corresponding to Fig. 3(c), the absorption efficiency of the metasurface grows up and then comes down, with the increase of the conductivity of VO$_2$. The highest absorption efficiency is about 98.9% at $\sigma = 8.6 \times 10^4 S/m$ (point A). The field pattern of point A is presented in the left of Fig. 3(b). It clearly shows that $(|E_{sca}|/|E_{in}|)^2$ is close to zero in the far field which verifies the perfect absorption property. The vertical white dashed line in Fig. 3(a) represents the absorption efficiency of the metasurface when $\sigma = 8.6 \times 10^4 S/m$, corresponding to Fig. 3(d). We find that nearly perfect absorption (>90%) is obtained for the incidence from $-34°$ to $34°$. Especially, with $\theta_i = 30°$ the absorption efficiency can reach 91.5% (point B). The corresponding field pattern of point B is shown in the right pattern of Fig. 3(b) and it clearly shows that only a little incident wave is reflected back. Therefore, by changing the conductivity of VO$_2$ from $\sigma = 200 S/m$ to $\sigma = 8.6 \times 10^4 S/m$, the functionality of the metasurface is switched from a nearly perfect retroreflector to a nearly perfect absorber at $\theta_i = \pm 30°$ and $\theta_i = 0°$. Notably, as the phase difference of binary unit cell is frequency-dependent, the proposed metasurface with good performance can only work in a narrow bandwidth near 5 THz.

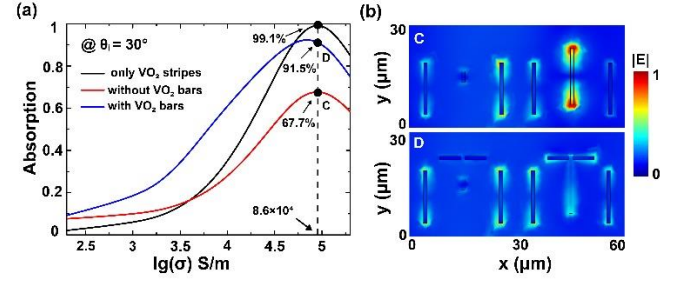

Fig. 4. (a) The absorption efficiency at $\theta_i = 30°$ with only VO$_2$ stripes (black line), without (red line) and with (blue line) gapped VO$_2$ bars. (b) Field distributions at the surface of z=8.0 μm for the metasurface without (upper) and with (lower) horizontal VO$_2$ bar at $\theta_i = 30°$.

In fact, this dual function is not a simple superposition of the retroreflection and absorption at the same frequency, but is a result of their interaction or interplay. To uncover it, we first consider a case that the Au stripes and the gapped VO$_2$ bars are removed from the binary metasurface in Fig. 2(a) at $\theta_i = 30°$. For the rest structure, Fig. 4(a) shows its absorption efficiency as a function of the conductivity of VO$_2$ (see the black line). Perfect absorption can be achieved when $\sigma = 8.6 \times 10^4 S/m$, as a result of strong resonance in lossy metallic VO$_2$ strips. Further, the Au stripes for designed retroreflector are added into the structure, which will decline dramatically the absorption efficiency (see the red line in Fig. 4(a)). In particular, when $\sigma = 8.6 \times 10^4 S/m$, absorption efficiency is about 66.7% (point C). The reason is that the added Au stripes contribute to the high-efficiency retroreflection. Fig. 4(b) shows electric field pattern at point C, where strong resonance occurs in Au stripes and VO$_2$ stripes, respectively. The resonance in Au produces high-efficiency retroreflection, while the resonance in VO$_2$ produces nearly perfect absorption. The declined absorption efficiency is a result of their interplay. Therefore, bifunctional metasurface in current case could not be realized by simply combining two functionalities together. The interaction between them is also need to take into consideration. To realize perfect absorption, we need to suppress the resonance of Au stripes. This is why additional horizontal gapped VO$_2$ bars are introduced. The corresponding absorption curve (blue curve) is shown in Fig. 4(a). It shows that the absorption efficiency of the metasurface with $\sigma = 8.6 \times 10^4 S/m$ increases to 91.5% (point D) with an increase of 23.8% compared to point C. The corresponding electric field pattern of point D is shown in Fig. 4(b) (lower field pattern). It could be clearly seen that the resonance of the long Au stripes is suppressed when they contact to the horizontal metallic VO$_2$ bars, because when the long Au stripes contact to the metallic VO$_2$ bars, the length of the long Au stripes equivalently increases, which makes the Au stripes deviate from its resonant length. The functionality of the finally designed metasurface is then determined by the competition between radiation loss and intrinsic loss of the whole system. When the metasurface is dominated by the radiation loss, it works as a retroreflector, and when the metasurface is dominated by the internal loss, it works as a nearly perfect absorber.

In conclusion, based on the phase transition property of $VO_2$, we have designed a binary metasurface with switchable functionality in THz regime. We numerically revealed that with the increase of temperature of $VO_2$, the functionality of the metasurface can well switch from three channel retroreflector to perfect absorber. At the low temperature, the metasurface works as multi-channel retroreflector with efficiency about 90.6%. In addition, quasi-retroreflector with efficiency beyond 85.0% can occur in a wide incident range. At the high temperature, as $VO_2$ turns into metal with high loss, the designed metasurface behaves as a perfect absorber at normal incidence. In addition, the absorption efficiency above 90.0% in the metasurface can occur within a wide incident range. By actively changing the temperature of $VO_2$, the switchable ability of the designed enables it visible and invisible, which could be potentially used for detection and anti-detection system. Considering recent advances in nanofabrication technology, metasurfaces consisting of structures of multiple materials [31, 32] have been experimentally realized, and therefore our proposed metasurface is feasible in sample fabrication. Although only theoretical result is displayed, it can facilitate more study on multifunctional metasurfaces with switchable effect, and we also expect the proposed device can be experimentally realized in the coming future. By selecting suitable phase-change materials or other tuning methods, similar ideas can be extended to other frequency range including microwave, infrared range and visible range, also to other wave dynamics, such as sound waves and water waves.

**Funding.** The National Natural Science Foundation of China (grant nos. 11974010, 11904169, 11604229, 11874311, 11774252); the Natural Science Foundation of Jiangsu Province (grant nos. BK20171206 and BK20190383); a project funded by the China Postdoctoral Science Foundation (grant no. 2018T110540); the Qing Lan project; the "333" project (BRA2015353); a project funded by State Key Laboratory of Metamaterial Electromagnetic Modulation Technology (grant no. GYL08-1458); the Priority Academic Program Development (PAPD) of Jiangsu Higher Education Institutions.